\documentstyle[amssymb,preprint,aps]{revtex}

\begin{document}
\draft
\title{ac Josephson effect in the resonant tunneling through mesoscopic
superconducting junctions}
\author{{\ Yu Zhu}$^1$,{\ Qing-feng Sun}$^2$, Wei Li$^1$, {Zhong-shui Ma}$^1${, and }%
Tsung-han Lin$^{1*}$}
\address{$^1${\it State Key Laboratory for Mesoscopic Physics and }\\
{\it Department of Physics, Peking University,}{\small \ }{\it Beijing}\\
{\it 100871, China}}
\address{$^2${\it Center for the Physics of Materials and}\\
{\it Department of Physics, McGill University, Montreal,}\\
{\it PQ, Canada H3A 2T8}}
\maketitle

\begin{abstract}
We investigate ac Josephson effect in the resonant tunneling through
mesoscopic superconducting junctions. In the presence of microwave
irradiation, we show that the trajectory of multiple Andreev reflections can
be closed by emitting or absorbing photons. Consequently, photon-assisted
Andreev states are formed and play the role of carrying supercurrent. On the
Shapiro steps, dc component appears when the resonant level $E_0$ is near
the position $V_L+\frac 12m_1\omega =V_R+\frac 12m_2\omega $ ($e=\hbar =1$),
where $m_1$ and $m_2$ are integers, $\omega $ is the frequency of microwave, 
$V_L$ and $V_R$ the chemical potentials of the superconducting leads.
Analytical result is derived in the limit $\Delta \rightarrow \infty $,
based on which new features of ac Josephson effect are revealed.
\end{abstract}



\baselineskip 20pt 
\newpage

{\it Introduction--}Since the achievement of atom-size superconducting
quantum point contact (SQPC) by break junction technique \cite
{SQPC1,SQPC2,SQPC3,SQPC4}, considerable theoretical works have been devoted
to the transport problem in S-I-S or S-N-S structures (S=superconductor,
I=tunnel barrier, N=normal metal). Comparison between experimental data and
theoretical calculation suggests that the coherent picture of multiple
Andreev reflection (MAR) is the central conception in understanding the
transport through such mesoscopic SQPC. Nevertheless, less attention is paid
to the resonant tunneling (especially in ac transport) through mesoscopic
superconducting junctions, in which the central region is consist of
discrete electronic states. Such resonant tunneling can be achieved in
nanoparticles sandwiched between superconducting membranes \cite{SDS1}, or
metallic clusters absorbed on the STM substrate \cite{SDS2}, or a piece of
carbon nanotube tunnel coupled to superconducting electrodes \cite{SDS3}, or
quantum dots fabricated in the S/2DEG hybrid structures \cite{SDS4}, etc.

The transport through these structures is greatly modified due to the
existence of discrete energy levels in the central region. It is shown by
Yeyati $et\;al.$ \cite{Yeyati} and Johansson $et\;al.$ \cite{Johansson} that
the combination of MAR processes and resonant transmission gives rise to a
rich subharmonic gap structure (SGS). Let us consider the transport through
a resonant level with width $\Gamma $ coupled to S\ leads with
superconducting gap $\Delta $. In the limit $\Gamma \gg \Delta $, the
resonance is sufficient broadened such that the I-V characteristics mimic
that of SQPC. In the regime $\Gamma \thicksim \Delta $, pronounced SGS
appears with concomitant negative differential conductance, which is
distinguished from the nonresonant transport through SQPC. In the limit $%
\Gamma \ll \Delta $, although SGS\ is more particular in the log-scale plot,
the tunnel current as a whole is exponentially small. This is because a
narrow resonance can not cover all the MAR\ trajectory, and it is very
unlikely that multiple discrete levels are just in the position where MAR\
trajectory passes (see Fig.1a).

In this paper, we address the effect of microwave (MW) irradiation on the
resonant tunneling through mesoscopic superconducting junctions. We show
that the tunnel current is greatly enhanced on the Shapiro steps, when the
resonant level $E_0$ is near the position $V_L+\frac 12m_1\omega =V_R+\frac 1%
2m_2\omega $ ($e=\hbar =1$ throughout), where $m_1$ and $m_2$ are integers, $%
\omega $ is the frequency of MW, $V_L$ and $V_R$ are the chemical potentials
of left and right S lead. This can be attributed to the formation of
photon-assisted Andreev states (PAAS), which play the role of carrying
supercurrent. In the limit $\Delta \rightarrow \infty $, analytical results
are derived, revealing new features of ac Josephson effect in the case of
resonant tunneling.

{\it Hamiltonian and Formulation--}The model Hamiltonian reads $%
H=\sum_{\beta =L,R}H_\beta +H_{cent}+H_T$, in which $H_\beta =\sum_{k\sigma
}\varepsilon _ka_{\beta k\sigma }^{\dagger }a_{\beta k\sigma }+\sum_k(\Delta
e^{\text{i}\phi _\beta }a_{\beta k\uparrow }^{\dagger }a_{\beta -k\downarrow
}^{\dagger }+H.c.)$ is the Hamiltonian for the $\beta $th S lead, $%
H_{cent}=E_0\sum_\sigma c_\sigma ^{\dagger }c_\sigma $ the resonant level in
the central region, and $H_T=\sum_{\beta k\sigma }(v_\beta \exp \left[ \text{%
i}\int_0^t\tilde{V}_\beta (t^{\prime })dt^{\prime }\right] a_{\beta k\sigma
}^{\dagger }c_\sigma +H.c.)$ the tunnel couplings. $\tilde{V}_\beta
(t)\equiv V_\beta +W_\beta \cos \omega t$ is the time-dependent voltage drop
across the $\beta $th barrier, where $V_\beta $ is the chemical potential
controlled by dc bias voltage and $W_\beta \cos \omega t$ the ac voltage
induced by MW irradiation. For simplicity, we assume that Coulomb
interaction and multi-level effect can be ignored in the central region.
This assumption is somehow too ideal to achieve, but allows us to obtain
analytical results and therefore instructive for the understanding of more
complicated cases.

Keldysh Green functions are defined in the $2\times 2$ Nambu representation: 
\begin{equation}
G_{t_1t_2}^{r,a,<}\equiv \left( 
\begin{array}{cc}
\langle \langle c_{\uparrow }(t_1)|c_{\uparrow }^{\dagger }(t_2)\rangle
\rangle ^{r,a,<} & \langle \langle c_{\uparrow }(t_1)|c_{\downarrow
}(t_2)\rangle \rangle ^{r,a,<} \\ 
\langle \langle c_{\downarrow }^{\dagger }(t_1)|c_{\uparrow }^{\dagger
}(t_2)\rangle \rangle ^{r,a,<} & \langle \langle c_{\downarrow }^{\dagger
}(t_1)|c_{\downarrow }(t_2)\rangle \rangle ^{r,a,<}
\end{array}
\right) ,
\end{equation}
The time-dependent current flowing out of the $\beta $th lead can be
expressed as 
\begin{equation}
I_\beta (t)=2%
\mathop{\rm Re}%
Tr\sigma _z(G^r\circ \Sigma _\beta ^{<}+G^{<}\circ \Sigma _\beta ^a)_{tt},
\end{equation}
in which $\sigma _z$ is the 3rd Pauli matrix, $\circ $ the shorthand
notation for integration over intermediate time variable, and $\Sigma _\beta 
$ the self energy due to tunnel coupling between the central region and the $%
\beta $th lead. $G^r$, $G^a$, and $G^{<}$ satisfy the integral equation: 
\begin{eqnarray}
G^{r,a} &=&g^{r,a}+g^{r,a}\circ \Sigma ^{r,a}\circ G^{r,a}, \\
G^{<} &=&G^r\circ \Sigma ^{<}\circ G^a,
\end{eqnarray}
in which 
\begin{eqnarray}
g_{t_1t_2}^{r,a} &=&\int \frac{d\epsilon }{2\pi }e^{-\text{i}\epsilon
(t_1-t_2)}g^{r,a}(\epsilon ), \\
\Sigma _{\beta ,t_1t_2}^{r,a,<} &=&U_\beta (t_1)\int \frac{d\epsilon }{2\pi }%
e^{-\text{i}\epsilon (t_1-t_2)}\Sigma _\beta ^{r,a,<}(\epsilon )U_\beta
^{\dagger }(t_2),
\end{eqnarray}
where 
\begin{eqnarray}
g^{r,a}(\epsilon ) &=&\left( 
\begin{array}{cc}
\frac 1{\epsilon -E_0\pm \text{i}0^{+}} & 0 \\ 
0 & \frac 1{\epsilon +E_0\pm \text{i}0^{+}}
\end{array}
\right) , \\
\Sigma _\beta ^{r,a}(\epsilon ) &=&\mp \frac{\text{i}}2\Gamma _\beta \frac{%
\epsilon \pm \text{i}\eta }{\sqrt{(\epsilon \pm \text{i}\eta )^2-\Delta ^2}}%
\left( 
\begin{array}{cc}
1 & -\frac \Delta {\epsilon \pm \text{i}\eta }e^{-\text{i}\phi _\beta } \\ 
-\frac \Delta {\epsilon \pm \text{i}\eta }e^{+\text{i}\phi _\beta } & 1
\end{array}
\right) , \\
\Sigma _\beta ^{<}(\epsilon ) &=&f(\epsilon )\left[ \Sigma _\beta
^a(\epsilon )-\Sigma _\beta ^r(\epsilon )\right] , \\
U_\beta (t) &=&\left( 
\begin{array}{cc}
\exp \left[ -\text{i}\int_0^t\tilde{V}_\beta (t^{\prime })dt^{\prime
}\right]  & 0 \\ 
0 & \exp \left[ +\text{i}\int_0^t\tilde{V}_\beta (t^{\prime })dt^{\prime
}\right] 
\end{array}
\right) ,
\end{eqnarray}
with $\Gamma _\beta $ being the coupling strength, $\eta $ the dephasing
rate in the S lead, $f(\epsilon )$ the Fermi distribution function, and $%
\mathop{\rm Im}%
\sqrt{z}>0$ as a convention. The remaining task is to solve these integral
equations and evaluate the dc component of the time-dependent current.

There are two intrinsic frequencies in the problem, $\omega _1=2V=2(V_L-V_R)$
and $\omega _2=\omega $. Generally, one may perform a Fourier transform 
\begin{equation}
A_{t_1t_2}=\sum_{l_1l_2}e^{\text{i}(l_1\omega _1+l_2\omega _2)t_1}\int \frac{%
d\epsilon }{2\pi }e^{-\text{i}\epsilon (t_1-t_2)}\tilde{A}_{l_1l_2}(\epsilon
),
\end{equation}
and derive the recursive relation for $\tilde{A}_{l_1l_2}(\epsilon )$, as
done in \cite{Cuevas}. The calculation in this way relies on the numerical
computing power, and analytical results are impossible. We note, however,
that in the case of narrow resonance, dc component appears only if $\omega
_1=N\omega _2$ with $N$ being an integer. This becomes clear by considering
PAAS shown in Fig.1b and 1c: an electron (a hole) incomes through the
resonant level $E_0$ and Andreev reflected by the right superconducting gap,
the reflected hole (the reflected electron) exchanges $m_1$ photons with the
MW\ field so that it can again passes through $E_0$. Then the hole (the
electron) is Andreev reflected by the right superconducting gap as an
electron (a hole), and exchanges $m_2$ photons to close the trajectory
(detailed discussion for photon-assisted Andreev reflection is available in 
\cite{ar2}). It is easy to see that the formation of PAAS requires $%
2V=(m_1+m_2)\omega $. For this reason, we shall only consider the case of $%
V=V_N=\frac N2\omega $, while the current deviated from this condition is
negligible small.

The problem is largely simplified since $\omega $ can be used as the basic
frequency in the Fourier expansion. Define the Fourier transformation as 
\begin{eqnarray}
A_{t_1t_2} &=&\sum_le^{\text{i}l\omega t_1}\int \frac{d\epsilon }{2\pi }e^{-%
\text{i}\epsilon (t_1-t_2)}\tilde{A}_l(\epsilon ), \\
{\bf A}_{mn}(\epsilon ) &=&\tilde{A}_{m-n}(\epsilon -n\omega ).
\end{eqnarray}
The definition guarantees the nice property that if $C=A\circ B$ then ${\bf C%
}_{mn}(\epsilon )=\sum_k{\bf A}_{mk}(\epsilon ){\bf B}_{kn}(\epsilon )$. The
Fourier transformed $g^{r,a}$ and $\Sigma _\beta ^{r,a,<}$ are 
\begin{eqnarray}
{\bf g}_{mn}^{r,a}(\epsilon ) &=&\delta _{mn}g^{r,a}(\epsilon -n\omega ), \\
{\bf \Sigma }_{R,mn}^{r,a,<}(\epsilon ) &=&\sum_l\left( 
\begin{array}{cc}
J_{l-m}(\alpha _R)\Sigma _{R,11}^{r,a,<}(\epsilon _l^0)J_{l-n}(\alpha _R) & 
J_{l-m}(\alpha _R)\Sigma _{R,12}^{r,a,<}(\epsilon _l^0)J_{n-l}(\alpha _R) \\ 
J_{m-l}(\alpha _R)\Sigma _{R,21}^{r,a,<}(\epsilon _l^0)J_{l-n}(\alpha _R) & 
J_{m-l}(\alpha _R)\Sigma _{R,22}^{r,a,<}(\epsilon _l^0)J_{n-l}(\alpha _R)
\end{array}
\right) , \\
{\bf \Sigma }_{L,mn}^{r,a,<}(\epsilon ) &=&\sum_l\left( 
\begin{array}{cc}
J_{l-m}(\alpha _L)\Sigma _{L,11}^{r,a,<}(\epsilon _l^{-})J_{l-n}(\alpha _L)
& J_{l-m-N}(\alpha _L)\Sigma _{L,12}^{r,a,<}(\epsilon _l^{+})J_{n-l}(\alpha
_L) \\ 
J_{m-l-N}(\alpha _L)\Sigma _{L,21}^{r,a,<}(\epsilon _l^{-})J_{l-n}(\alpha _L)
& J_{m-l}(\alpha _L)\Sigma _{L,22}^{r,a,<}(\epsilon _l^{+})J_{n-l}(\alpha _L)
\end{array}
\right) ,
\end{eqnarray}
in which $\alpha _\beta \equiv \frac{W_\beta }\omega $ is the MW\ strength
on the $\beta $th tunnel barrier, $J_n(x)$ the $n$th Bessel function, $%
\epsilon _l^0\equiv \epsilon -l\omega $ and $\epsilon _l^{\pm }\equiv
\epsilon -l\omega \pm V_N$, $V_L=V_N$ and $V_R=0$ are set as a convention.
Correspondingly, the equations for $G^r$, $G^a$, and $G^{<}$ are Fourier
transformed into ${\bf G}^{r,a}(\epsilon )={\bf g}^{r,a}(\epsilon )+{\bf g}%
^{r,a}(\epsilon ){\bf \Sigma }^{r,a}(\epsilon ){\bf G}^{r,a}(\epsilon )$ and 
${\bf G}^{<}(\epsilon )={\bf G}^r(\epsilon ){\bf \Sigma }^{<}(\epsilon ){\bf %
G}^a(\epsilon )$, or equivalently, 
\begin{eqnarray}
{\bf G}^{r,a}(\epsilon ) &=&{\bf g}^{r,a}(\epsilon )+{\bf g}^{r,a}(\epsilon )%
{\bf \Sigma }^{r,a}(\epsilon ){\bf g}^{r,a}(\epsilon )+{\bf g}%
^{r,a}(\epsilon ){\bf \Sigma }^{r,a}(\epsilon ){\bf g}^{r,a}(\epsilon ){\bf %
\Sigma }^{r,a}(\epsilon ){\bf g}^{r,a}(\epsilon )+\cdots , \\
{\bf G}^{<}(\epsilon ) &=&\left[ {\bf g}^r(\epsilon )+{\bf g}^r(\epsilon )%
{\bf \Sigma }^r(\epsilon ){\bf g}^r(\epsilon )+\cdots \right] {\bf \Sigma }%
^{<}(\epsilon )\left[ {\bf g}^a(\epsilon )+{\bf g}^a(\epsilon ){\bf \Sigma }%
^a(\epsilon ){\bf g}^a(\epsilon )+\cdots \right] .
\end{eqnarray}

We note that finite order perturbation expansion is inadequate in the
problem, because the formation of PAAS\ involves up to infinite order of
tunneling processes. To re-sum up the series, we adopt the resonant
approximation \cite{ar2} 
\begin{equation}
\frac 1{(\epsilon -l_1\omega \pm E_0+\text{i}0^{+})}\cdot \frac 1{(\epsilon
-l_2\omega \pm E_0+\text{i}0^{+})}\thickapprox \delta _{l_1l_2}\frac 1{%
(\epsilon -l_1\omega \pm E_0+\text{i}0^{+})^2}\;.
\end{equation}
The approximation implies that the overlap between sidebands $E_0+l_1\omega $
and $E_0+l_2\omega $ can be ignored if $l_1\neq l_2$, which is justified
when $\Gamma _\beta \ll \omega $. Applying the approximation to Eq.(17) and
Eq.(18), one can obtain the solution 
\begin{equation}
{\bf G}_{mn}^{r,a,<}(\epsilon )=\left( 
\begin{array}{cc}
\delta _{mn}\sum_k\widehat{G}_{mk,11}^{r,a,<} & \widehat{G}_{mn,12}^{r,a,<}
\\ 
\widehat{G}_{nm,21}^{r,a,<} & \delta _{mn}\sum_k\widehat{G}_{kn,22}^{r,a,<}
\end{array}
\right) ,
\end{equation}
in which 
\begin{eqnarray}
\widehat{G}_{mn}^{r,a} &=&\left[ \left( \widehat{g}_{mn}^{r,a}\right) ^{-1}-%
\widehat{\Sigma }_{mn}^{r,a}\right] ^{-1}, \\
\widehat{G}_{mn}^{<} &=&\widehat{G}_{mn}^r\widehat{\Sigma }_{mn}^{<}\widehat{%
G}_{mn}^a\;,
\end{eqnarray}
with 
\begin{eqnarray}
\widehat{g}_{mn}^{r,a} &=&\left( 
\begin{array}{cc}
{\bf g}_{mm,11}^{r,a}(\epsilon ) & 0 \\ 
0 & {\bf g}_{nn,22}^{r,a}(\epsilon )
\end{array}
\right) , \\
\widehat{\Sigma }_{mn}^{r,a,<} &=&\left( 
\begin{array}{cc}
{\bf \Sigma }_{mm,11}^{r,a,<}(\epsilon ) & {\bf \Sigma }_{mn,12}^{r,a,<}(%
\epsilon ) \\ 
{\bf \Sigma }_{nm,21}^{r,a,<}(\epsilon ) & {\bf \Sigma }_{nn,22}^{r,a,<}(%
\epsilon )
\end{array}
\right) .
\end{eqnarray}
It can be shown that the relations ${\bf G}^r=\left( {\bf G}^a\right)
^{\dagger }$ and ${\bf G}^{<}=-\left( {\bf G}^{<}\right) ^{\dagger }$ still
hold in the solution.

With these Green functions, the time-dependent current can be expressed in a
summation over ac components 
\begin{equation}
I_\beta (t)=2%
\mathop{\rm Re}%
\sum_le^{\text{i}l\omega t}\left\{ \int \frac{d\epsilon }{2\pi }Tr\sigma
_z\left[ {\bf G}^r(\epsilon ){\bf \Sigma }_\beta ^{<}(\epsilon )+{\bf G}%
^{<}(\epsilon ){\bf \Sigma }_\beta ^a(\epsilon )\right] _{l0}\right\} \;.
\end{equation}
The current formula can be applied to the study of ac harmonics. However, we
are more interested in the dc component $\bar{I}=\bar{I}_L=-\bar{I}_R$, due
to the experimental reasons. To produce analytical results, we take the
limit $\Delta \rightarrow \infty $, due to which all single particle
processes are forbidden and Andreev reflection is the only conducting
mechanism. After some algebra and take the limit $\eta \rightarrow 0$
(notice that $\lim_{\eta \rightarrow 0}\frac \eta {x^2+\eta ^2}=\pi \delta
(x)$ ), one can derive 
\begin{eqnarray}
\bar{I} &=&2\sin \phi \frac{\Gamma _L\Gamma _R}4(-)\sum_mJ_{m-N}(2\alpha
_L)J_m(2\alpha _R)\frac 1{\left( E_m^{+}-E_m^{-}\right) ^2}\left[
F(E_m^{+})+F(E_m^{-})\right] , \\
F(\epsilon ) &=&\left( \epsilon +E_0-\frac 12m\omega \right) \sum_l\left[
f\left( \epsilon _l^0+\frac 12m\omega \right) J_l^2(\alpha _R)+f\left(
\epsilon _l^{-}+\frac 12m\omega \right) J_l^2(\alpha _L)\right] + \\
&&\left( \epsilon -E_0+\frac 12m\omega \right) \sum_l\left[ f\left( \epsilon
_l^0-\frac 12m\omega \right) J_{-l}^2(\alpha _R)+f\left( \epsilon _l^{+}-%
\frac 12m\omega \right) J_{-l}^2(\alpha _L)\right] \;.  \nonumber
\end{eqnarray}
in which $E_m^{\pm }=\pm \sqrt{\left( E_0-\frac 12m\omega \right) ^2+\left|
\Gamma _m\right| ^2}$ and $\Gamma _m=\frac 12\left[ \Gamma _RJ_m(2\alpha
_R)+\Gamma _Le^{-\text{i}\phi }J_{m-N}(2\alpha _L)\right] $. Eq.(26) is the
central result of this paper, which is for the dc component of Josephson
current in the resonant tunneling through mesoscopic superconducting
junctions.

Before numerical study, we make a few remarks on this result: (1) The phase
dependence of $\bar{I}$ is mainly determined by the prefactor $\sin \phi $,
and a weak $\cos \phi $ dependence is hidden in $E_m^{\pm }$ via $\Gamma _m$%
. For this reason, $\phi $ is set as $\frac \pi 2$ in the numerical
calculation. (2) For the special case $N=0$, $\alpha _L=\alpha _R=0$, and $%
\Gamma _L=\Gamma _R=\Gamma $, $\bar{I}=2\sin \phi \frac{\Gamma ^2}4\frac 1{%
\tilde{E}_0}\left[ f\left( -\tilde{E}_0\right) -f\left( +\tilde{E}_0\right)
\right] $, with $\tilde{E}_0=\sqrt{E_0^2+\left( \Gamma \cos \frac \phi 2%
\right) ^2}$, which reproduces the exact result for dc Josephson current
through a resonant level. (3) The current formula is for the gauge choice of 
$V_L=V_N$ and $V_R=0$. It is easy to see that the formula is invariant under
the transformation $L\longleftrightarrow R$, $V_N\rightarrow -V_N$, $%
N\rightarrow -N$, and $E_0\rightarrow E_0-V_N$, which corresponds to the
gauge choice $V_L=0$ and $V_R=-V_N$. (4) $E_m^{\pm }$ are the poles of ${\bf %
G}^r(\epsilon )$, which can be interpreted as PAAS. Obviously, $E_m^{\pm }$
contribute to the supercurrent with opposite signs. One can expect that
resonant structures will appear near $E_0=\frac 12m\omega $. (5) Bessel
functions enter the current formula not in the square form $J_n^2(x)$. It
can be shown that $\bar{I}(\alpha _L,\alpha _R,\phi )=$ $\bar{I}(-\alpha
_L,-\alpha _R,\phi )$ for $N$ even and $\bar{I}(\alpha _L,\alpha _R,\phi )=$ 
$\bar{I}(-\alpha _L,-\alpha _R,\phi +\pi )$ for $N$ odd.

{\it numerical results}--Firstly, we discuss the case that MW\ is applied
symmetrically to the left and right tunnel barrier, i.e., $\alpha _L=\alpha
_R$. Fig.2 shows the curves of $I_c=\bar{I}(\phi =\frac \pi 2)$ vs $E_0-%
\frac{V_N}2$ at bias voltage $V_N=\frac N2\omega $, with $N=0,1,2,3,4$ (the
curves are shifted by $\frac{V_N}2$ for demonstration). Four features are
noteworthy in the plot: (1) The curve is symmetric (anti-symmetric) with
respect to $E_0=\frac{V_L+V_R}2=\frac{V_N}2$ for $N$ even (odd). (2)
Photon-assisted structures appear near $E_0=V_L-\frac 12(N-m)\omega =V_R+%
\frac 12m\omega $. (3) There are two types of resonant structures, single
peak (dip) and peak-dip pair. (4) The structures grow with the MW strength
non-monotonously.. Feature (1) is due to the relation $I(E_0,\phi
)=I(V_N-E_0,\phi )$ for $N$ even and $I(E_0,\phi )=I(V_N-E_0,\phi +\pi )$
for $N$ odd. Using $I(E_0,-\phi )=-I(E_0,\phi )$, one can obtain $I(E_0,\phi
=\frac \pi 2)=(-)^NI(V_N-E_0,\phi =\frac \pi 2)$. Feature (2) can be
understood in terms of PAAS shown in Fig.1b and 1c: Electron and hole are
Andreev reflected back and forth by the superconducting gaps. When $E_0$ is
near the position $V_L-\frac 12(N-m)\omega =V_R+\frac 12m\omega $,
quasiparticles may exchange $N-m$ photons with the MW\ field at the left
tunnel barrier and exchange $m$ photons at the right, so that the trajectory
is closed and bound states are formed. It is these PAAS that play the role
of carrying supercurrent. Understanding of feature (3) and (4) needs
quantitative analysis of each photon-assisted structure. To proceed, we
decompose the total current into $\bar{I}=\sum_mI_m$, and expand $I_m$ near $%
E_0=\frac 12m\omega $. Let $E_0=\frac 12m\omega +\delta $, one can obtain $%
I_m=I_m^{+}+I_m^{-}$ with 
\begin{eqnarray}
I_m^{\pm } &=&2\sin \phi \frac{\Gamma ^2}4(-)J_{m-N}(2\alpha _L)J_m(2\alpha
_R)\frac 1{4\left( \delta ^2+\left| \Gamma _m\right| ^2\right) }  \nonumber
\\
&&\left[ \alpha _m^{\pm }\left( \pm \sqrt{\delta ^2+\left| \Gamma _m\right|
^2}+\delta \right) +\beta _m^{\pm }\left( \pm \sqrt{\delta ^2+\left| \Gamma
_m\right| ^2}-\delta \right) \right] .
\end{eqnarray}
The cancelation between $I_m^{+}$ and $I_m^{-}$ results in two types of
resonant structures: $C_1\frac 1{\sqrt{\delta ^2+\left| \Gamma _m\right| ^2}}%
+C_2\frac \delta {\delta ^2+\left| \Gamma _m\right| ^2}$, corresponding to
single peak (dip) and peak-dip pair. At zero temperature, $C_1$ and $C_2$
are proportional to 
\begin{eqnarray}
C_1 &\propto &\sum_l\left[ \delta _{l,\frac 12m}J_l^2(\alpha _R)+\delta _{l,%
\frac 12(m-N)}J_l^2(\alpha _L)\right] , \\
C_2 &\propto &\left[ \sum_{l>-\frac 12m}J_l^2(\alpha _R)-\sum_{l>\frac 12%
m}J_l^2(\alpha _R)+\sum_{l>-\frac 12(m-N)}J_l^2(\alpha _L)-\sum_{l>\frac 12%
(m-N)}J_l^2(\alpha _L)\right] \;,
\end{eqnarray}
The oscillatory nature of Bessel functions in these coefficients is
responsible for feature (4). There is an interesting special case, $N=2$, $%
m=1$, $\alpha _L=\alpha _R$, due to which $C_1=C_2=0$ for arbitrary MW
strengths. Correspondingly, the resonant structure near $E_0-\frac{V_N}2=0$
is always missing for $N=2$.

Next, we investigate the case that MW\ is applied only to one of the tunnel
barriers, i.e., $\alpha _L=0$ and $\alpha _R\neq 0$. Fig.3 shows the curves
of $I_c$ vs $E_0$ for $\alpha _L=0$ and $\alpha _R=1$. In contrast to the
symmetric case, a single peak is pinned at $E_0=V_L$. The reason is as
follows: since no MW is applied to the left barrier, Andreev tunneling
through this barrier occurs only when $E_0$ lines up with $V_L$, while
photon-assisted processes is allowed at the right barrier irradiated by MW\
field. The inset shows $I_c$ vs MW\ strength $\alpha _R$ at $E_0=V_L$ for $%
N>0$ (The case of $N<0$ can be easily deduced from $N>0$). One can see in
the plot that the peak height (including the sign) is predominated by the
prefactor $J_{m-N}(2\alpha _L)J_m(2\alpha _R)$. For $\alpha _L=0$, $%
J_{m-N}(2\alpha _L)$ requires $m=N$, and the peak height is proportionate to 
$J_N(2\alpha _R)$. We note that there exist some regions of MW\ strength
where $I_c$ and $V_N$ have opposite signs, which can be viewed as quantum
pump effect. However, this feature is dramatically different from the
quantum pump effect in normal mesoscopic junctions.

{\it conclusion--}To sum up, we have investigated ac Josephson effect in the
resonant tunneling through mesoscopic superconducting junctions. We show
that PAAS play an essential role for conducting supercurrent through a
narrow resonance, in which MAR\ trajectory is closed by exchanging photons
with MW\ field. On the Shapiro steps $2V=N\omega $, dc component appears
when the resonant level $E_0$ is near $V_L-(N-m)\omega =V_R+m\omega $, due
to the formation of PAAS. In the limit $\Delta \rightarrow \infty $,
analytical result, Eq.(26), is derived for the dc component of Josephson
current, and help to understand new features of ac Josephson effect in the
case of resonant tunneling. We have dropped Coulomb interaction in the
calculation, the derived results are meaningful in the following senses: (1)
They are directly applicable to the systems where Coulomb blockade effect is
negligible, i.e., $U\thicksim \Gamma \ll \Delta $ ($U$ is the charging
energy). This is possible by using substrate with large dielectric constant
to reduce $U$ or using proper material as S lead to obtain large $\Delta $.
(2) They are instructive for more complicated cases. We note that the
conception of PAAS is also useful for the interacting case. For $U\gg \Gamma 
$, the resonant level $E_0$ is effectively split into two resonances $E_0$
and $E_0+U$. Similar to Fig.1b and 1c, one can draw diagrams of closed MAR
trajectory through these resonances. Moreover, Coulomb blockade can be
partially removed by bias voltage or MW irradiation, as long as $U$ is
comparable to $\Delta $. Obviously, calculation including interaction term
is much more difficult, and analytical results are almost impossible.
Efforts along this line are still in progress.

This project was supported by NSFC\ under Grants No. 10074001 and No.
90103027, and also by the support from the Visiting Scholar Foundation of
State Key Laboratory for Mesoscopic Physics in Peking University.

\smallskip $^{*}$ To whom correspondence should be addressed.


\section*{Figure Captions}

\begin{itemize}
\item[{\bf Fig. 1}]  Schematic diagram of the resonant tunneling through
mesoscopic superconducting junctions. (a): Without MW\ irradiation, MAR\
trajectory can not be covered by a narrow resonance, and the tunnel current
is exponentially small. (b) and (c): In the presence of MW\ irradiation, MAR
trajectory can be closed by emitting or absorbing photons. Two types of
PAAS\ are formed, carrying supercurrent with opposite signs.

\item[{\bf Fig. 2}]  The curves of $I_c\equiv \bar{I}(\phi =\frac \pi 2)$ vs 
$E_0-\frac{V_N}2$ at bias voltage $V_N=\frac N2\omega $ ($N$ from $0$ to $4$%
), with symmetric MW\ strengths $\alpha _L=\alpha _R$. Parameters are: $%
\omega =1$, $\Gamma _L=\Gamma _R=0.02$, $k_BT=0.001$. $V_L=V_N$ and $V_R=0$
are set as a convention.

\item[{\bf Fig. 3}]  The curves of $I_c\equiv \bar{I}(\phi =\frac \pi 2)$ vs 
$E_0$ at bias voltage $V_N=\frac N2\omega $ ($N$ from $-4$ to $4$), with MW\
strengths $\alpha _L=0$ and $\alpha _R=1$. Other parameters are the same as
Fig.2. The inset shows $I_c$ at $E_0=V_L$ vs the MW strength $\alpha _R$.
Diagrams for corresponding PAAS is also shown in the plot.
\end{itemize}

\end{document}